# A Rubber-Modified Thermoplastic where the Morphology Produced by Phase-Separation Induced by Polymerization Disappears at High Conversions

E.R. Soulé, G.E. Eliçabe, R.J.J. Williams*

*Institute of Materials Science and Technology (INTEMA), University of Mar del Plata and National Research Council (CONICET), J. B. Justo 4302, 7600 Mar del Plata, Argentina*

* Corresponding author. Tel: 54 223 481 6600; fax: 54 223 481 0046.
*E-mail address:* williams@fi.mdp.edu.ar (R.J.J. Williams).

## Abstract

An unexpected experimental finding is reported where the primary morphology developed during polymerization-induced phase separation in a rubber-modified thermoplastic disappears at high conversions. This process was evidenced by light scattering (LS) and scanning electron microscopy (SEM) for a particular composition of solutions of polyisobutylene oligomers (PIB) in isobornylmethacrylate (IBoMA), during the free-radical polymerization of the monomer. The primary phase separation produced a dispersion of domains rich in PIB containing significant amounts of the monomer (IBoMA). Polymerization of the monomer in these domains occurred at high overall conversions producing the filling of dispersed domains with a PIBoMA-PIB blend. Under these conditions the final material had the appearance of a homogeneous blend. The process might be adapted to produce new types of rubber-modified thermoplastics where rubber particles are replaced by rubber-rich domains that do not exhibit definite boundaries.

## 1. Introduction

A phase separation process might take place when a polymerization is carried out employing a solution of a suitable modifier (e.g., rubber, thermoplastic polymer, liquid crystal) in the starting monomers. Thermodynamic factors that drive the polymerization-induced phase separation have been extensively discussed in the literature [1-5]. The main factor is the increase in the average size of reaction products and the corresponding decrease in the entropy of mixing. Secondary factors are the variation of the interaction parameter as a result of changes in the chemical structures, and elastic effects when a network is formed. Morphologies generated by this process depend on the initial composition and on the competition between the kinetics of polymerization and phase separation.

We have recently analyzed the phase separation process that occurs in solutions of polyisobutylene oligomers (PIB) in isobornylmethacrylate (IBoMA), during the free-radical polymerization of the monomer [6]. Under the conditions investigated, a dispersion of PIB-rich droplets was generated during polymerization. The average size of these domains increased with conversion attaining a plateau value. Then, we decided to investigate the effect of a prolonged heating at the polymerization temperature on the morphologies generated. In most cases no significant variation of morphologies was observed. But one set of blends exhibited a completely unexpected behaviour: at the end of the heating dispersed domains had almost disappeared and the material resembled a homogeneous blend. This curious behaviour has not been previously reported in the literature and will be analyzed in the present communication.

## 2. Experimental

The polymerization of a solution containing 15 wt % polyisobutylene (PIB, YPF-Repsol, $M_n$ = 1339 g/mol, $M_w$ = 3035 g/mol) in isobornyl methacrylate (IBoMA, Aldrich), was performed at 80 ºC in the presence of benzoyl peroxide (BPO, Akzo-Nobel, 2 wt % in the mixture with IBoMA).

Polymerization-induced phase separation was followed in situ by light scattering (LS) using an experimental set up described elsewhere [6]. In order to transform polymerization times into monomer conversions we used conversion vs. time curves obtained for by differential scanning calorimetry operating in the isothermal mode (DSC, Pyris 1, Perkin-Elmer). The main problem was the appearance of an induction time related to the consumption of the inhibitors initially present in the formulation. The duration of this induction period depends on variables that are very difficult to reproduce such as the temperature history and the concentration of dissolved oxygen. To circumvent this problem we fixed a point in the monomer conversion scale using the cloud-point conversion determined independently. Then the time scale was transformed into a monomer conversion scale using the polymerization kinetics. This procedure was used up to a conversion equal to 0.85, where the polymerization appeared to be practically arrested as inferred from the return of the isothermal DSC signal to the baseline. The slow increase in conversion after this point was followed measuring the evolution of the residual reaction heat in DSC scans (10 ºC/min).

SEM micrographs of fracture surfaces coated with a fine gold layer were obtained using a Jeol JSM 6460 LV device.

## 3. Results and Discussion

Fig. 1a shows LS spectra in the 0.506 – 0.788 conversion range while Fig. 1b shows LS spectra in the 0.801 – 0.93 conversion range. Starting at the cloud-point conversion ($p_{cp}$ = 0.41), a peak in the intensity (*I*) vs. scattering vector (*q*) spectra was generated with a maximum increasing with conversion up to conversion values close to 0.85 (Fig. 1 and 2). However, for conversions higher than 0.85 the intensity of the peak began to decrease and practically disappeared for *p* = 0.93 (Fig. 1b). The increasing period of the LS signal took place during about 4 min while the decreasing period was gradually observed during the following 130 min at 80 ºC (in a time scale the intensity of the peak increased at a fast rate and decreased at a slow rate).

The disappearance of the scattering peak in systems undergoing a polymerization-induced phase separation has already been reported in the literature and ascribed to the matching of refractive indices of both phases [7], or to a broadening of the distribution of dispersed particles [8]. We will show that in our case the decrease in the scattered intensity is associated to the disappearance of the morphology initially formed.

Particle-size distributions were obtained from LS spectra shown in Fig. 1, using a model proposed by Pedersen [9], as described in a previous publication [6]. Full curves shown in Fig. 1 indicate the excellent fitting of *I* vs. *q* curves obtained with this model. The evolution of the predicted average radius of dispersed-phase domains as a function of conversion is shown in Fig. 2. The average radius of the population increased with conversion attaining a limiting value at a conversion close to 0.85. A sharp decrease in the average radius was predictedby applying Pedersen's model in the 0.85 – 0.90 conversion range.

SEM images were obtained for samples kept at 80 ºC during different times in the decreasing period of the LS signal. A set of these images is shown in Fig. 3 for different overall conversions. A visual inspection shows that the concentration and size of dispersed-phase domains decreased with conversion. The average radius and volume fraction of dispersed-phase domains were estimated from SEM images following usual procedures [10]. The evolution of the average radius is plotted in Fig. 2 showing a reasonable agreement with LS data up to conversions where reliable information could be obtained from the latter technique. The decrease in the volume fraction of dispersed phase is plotted in Fig. 4.

Isothermal DSC scans at 80 ºC showed a sharp decrease of the signal, proportional to the polymerization rate, at conversions close to 0.80. At this conversion vitrification of the matrix took place producing a significant deceleration of the reaction rate. At conversions close to 0.85 the signal returned to the baseline. After this period, the overall conversion was calculated from the residual reaction heat compared to the value determined for the complete polymerization of the monomer [11]. As shown in Fig. 5, the residual heat was composed of two well-defined peaks. While the first peak decreased significantly with conversion the second one experienced only a small decrease. With this evidence, the first peak was assigned to the polymerization of residual monomer present in the dispersed phase, and the second peak was ascribed to the polymerization of residual monomer present in the matrix after its devitrification at about 120 ºC. Therefore, the sharp decrease of the average size of dispersed-phase domains observed at high conversions can be correlated to the continuation of polymerization in the dispersed phase. There is, however, a problem in this interpretation related to the fact that the fraction of polymerized monomer that corresponds to the first peak is higher than the maximum volume fraction of dispersed domains even with the (incorrect) assumption that these domains contain only monomer. The presence of an interphase between the glassy matrix and dispersed-phase droplets can reconcile the proposed explanation with the experimental observations.

Fig. 6 is a qualitative plot of concentration profiles in the region close to a droplet at the time when the matrix undergoes vitrification. The matrix is mainly composed of the polymerization product (PIBoMA) and small amounts of residual monomer (IBoMA) and modifier (PIB). Droplets are composed of a solution of PIB and IBoMA that were phase separated during polymerization. A thermodynamic analysis of the system showed that the phase separated at the cloud point was a binary blend of IBoMA and PIB containing 30 wt % IBoMA (an insignificant amount of PIBoMA was also present) [6]. Between both regions there is an interphase where gradients in the concentrations of the three components are established. After vitrification of the matrix the interphase and droplets remain as viscous liquids enabling polymerization of the residual monomer. Thermodynamics predicts a secondary phase separation of a PIBoMA-rich phase inside the droplets. However, the high viscosity of the blend imposes diffusional restrictions to this secondary phase separation and the result is that the originally dispersed domains become filled with a PIBoMA-PIB blend that could not be distinguished from the matrix employing either LS or SEM techniques. The result is the disappearance of the morphology of dispersed domains.

## 4. Conclusions

The primary morphology developed during a polymerization-induced phase separation in a rubber-modified thermoplastic can disappear for particular compositions when the polymer generated inside dispersed domains does not phase separate from the rubber due to diffusional restrictions. Under these conditions the contours of the initial dispersed domains are erased and the final material has the appearance of a homogeneous blend. In fact, the material is heterogeneous because the composition of the regions where dispersed domains were originally present must be different than the composition of the matrix. It would be interesting to adapt this procedure to synthesize and investigate properties of rubber-modified thermoplastics where the rubber-rich domains do not exhibit definite boundaries.

## Acknowledgment

We acknowledge the financial support of the following institutions of Argentina: University of Mar del Plata, National Research Council (CONICET), and National Agency for the Promotion of Science and Technology (ANPCyT).

**References**

[1] Williams RJJ, Borrajo J, Adabbo HE, Rojas AJ. In: Riew CK, Gillham JK, editors. Rubber-modified thermoset resins, Adv Chem Ser 208. American Chemical Society: Washington DC; 1984. p. 195-213. Chapter 13.

[2] Inoue T. Prog Polym Sci 1995;20:119.

[3] Clarke N, McLeish TCB, Jenkins SD. Macromolecules 1995;28:4650.

[4] Williams RJJ, Rozenberg BA, Pascault JP. Adv Polym Sci 1997;128:95.

[5] Eliçabe GE, Larrondo HA, Williams RJJ. Macromolecules 1998;31:8173.

[6] Soulé ER, Eliçabe GE, Borrajo J, Williams RJJ. Ind Eng Chem Res 2007, in press (web release date: 06-Mar-2007; DOI: 10.1021/ie061567c).

[7] Kim BS, Chiba T, Inoue T. Polymer 1995;36:67.

[8] Yang Y, Fujiwara H, Chiba T, Inoue T. Polymer 1998;39:2745.

[9] Pedersen JS. J App Cryst 1994;27:595.

[10] Verchère D, Pascault JP, Sautereau H, Moschiar SM, Riccardi CC, Williams, RJJ. J Appl Polym Sci 1991;42:701.

[11] Soulé ER, Borrajo J, Williams RJJ. Macromolecules 2005;38:5987.

## Legends to the Figures

Fig. 1. Light scattering spectra plotted as intensity (arbitrary units) as a function of the modulus of the scattering vector *q*. Points are experimental results and the curves represent the fitting obtained with Pedersen's model; (a) the intensity increases with monomer conversion (0.506, 0.548, 0.585, 0.617, 0,673, 0.719, 0.757, 0,788); (b) the intensity decreases with monomer conversion (0.801, 0.854, 0.856, 0.858, 0.859, 0.861, 0.865, 0.88, 0.93).

Fig. 2. Average radius of dispersed-phase droplets as a function of the monomer conversion (full symbols are values predicted from the fitting of LS curves using Pedersen's model; crosses are values determined in SEM images).

Fig. 3. SEM micrographs obtained at different monomer conversions; (a) 0.85, (b) 0.905, (c) 0.95.

Fig. 4. Evolution of the volume fraction of dispersed phase as a function of conversion.

Fig. 5. Polymerization of the residual monomer as shown by DSC scans of samples previously reacted to the indicated conversions.

Fig. 6. Qualitative plot of concentration profiles in a region close to a droplet at the time when the matrix vitrifies.

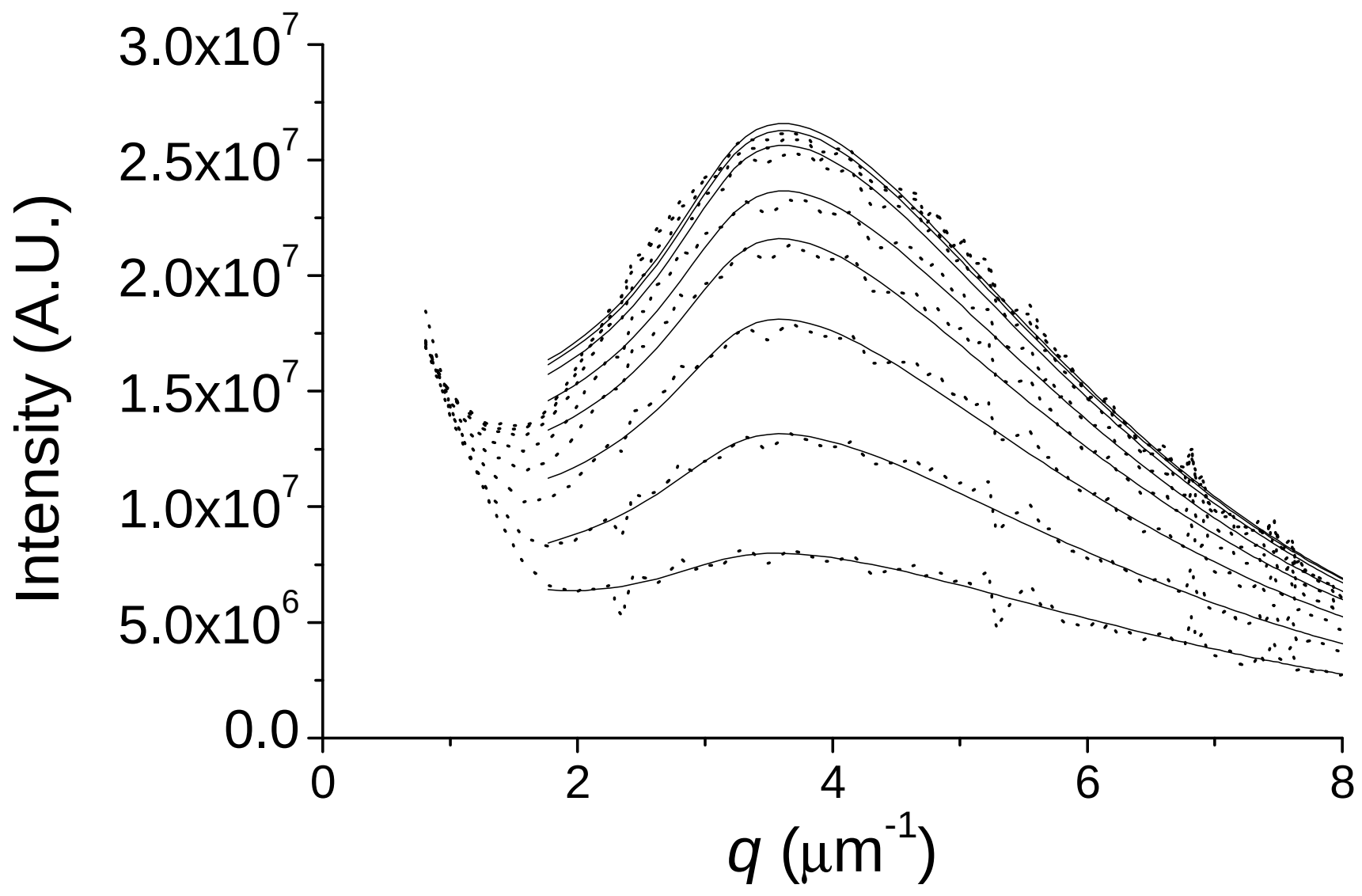

**(a)**

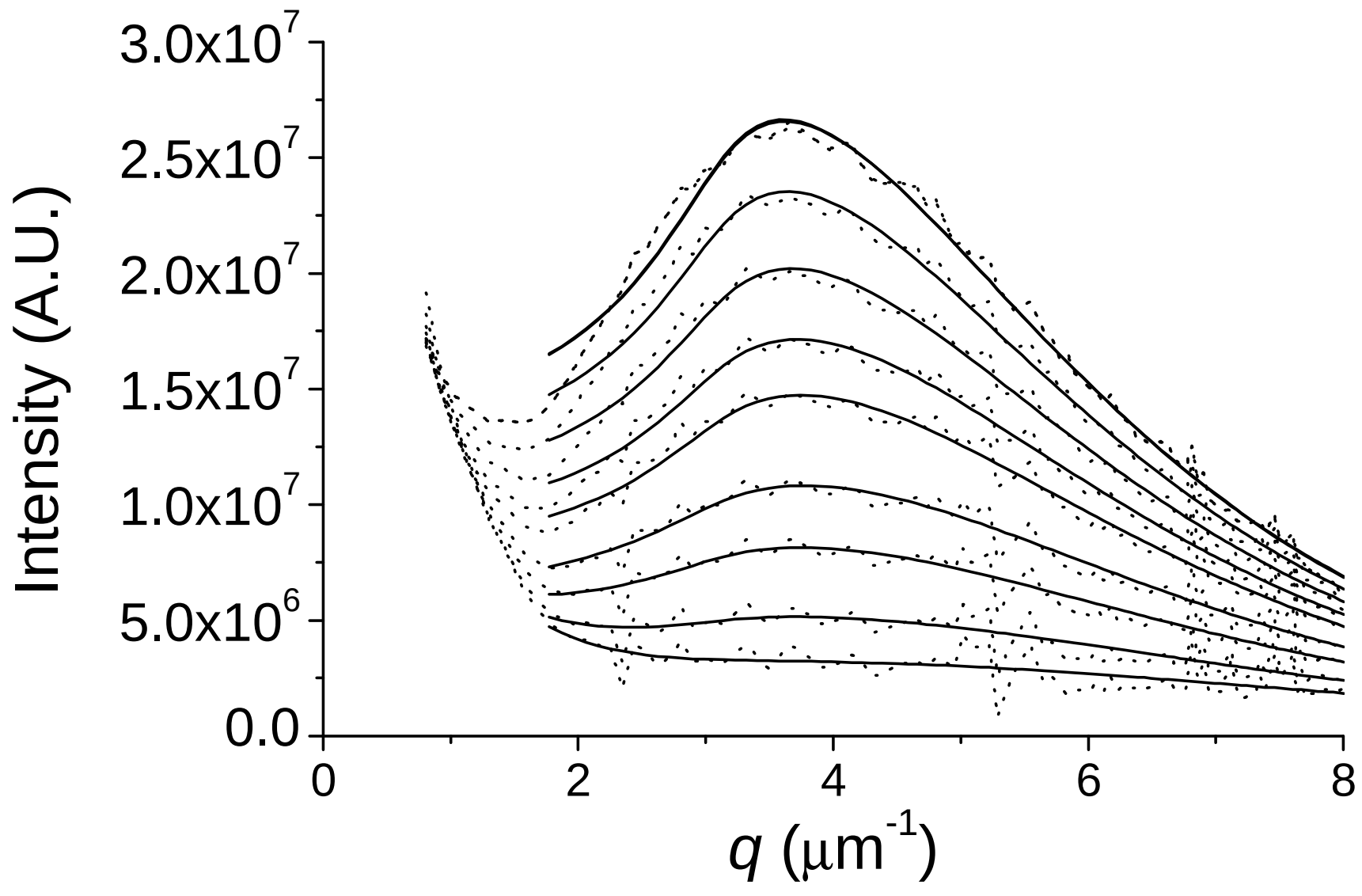

**(b)**

Fig. 1. Light scattering spectra plotted as intensity (arbitrary units) as a function of the modulus of the scattering vector *q*. Points are experimental results and the curves represent the fitting obtained with Pedersen's model; (a) the intensity increases with monomer conversion (0.506, 0.548, 0.585, 0.617, 0,673, 0.719, 0.757, 0,788); (b) the intensity decreases with monomer conversion (0.801, 0.854, 0.856, 0.858, 0.859, 0.861, 0.865, 0.88, 0.93).

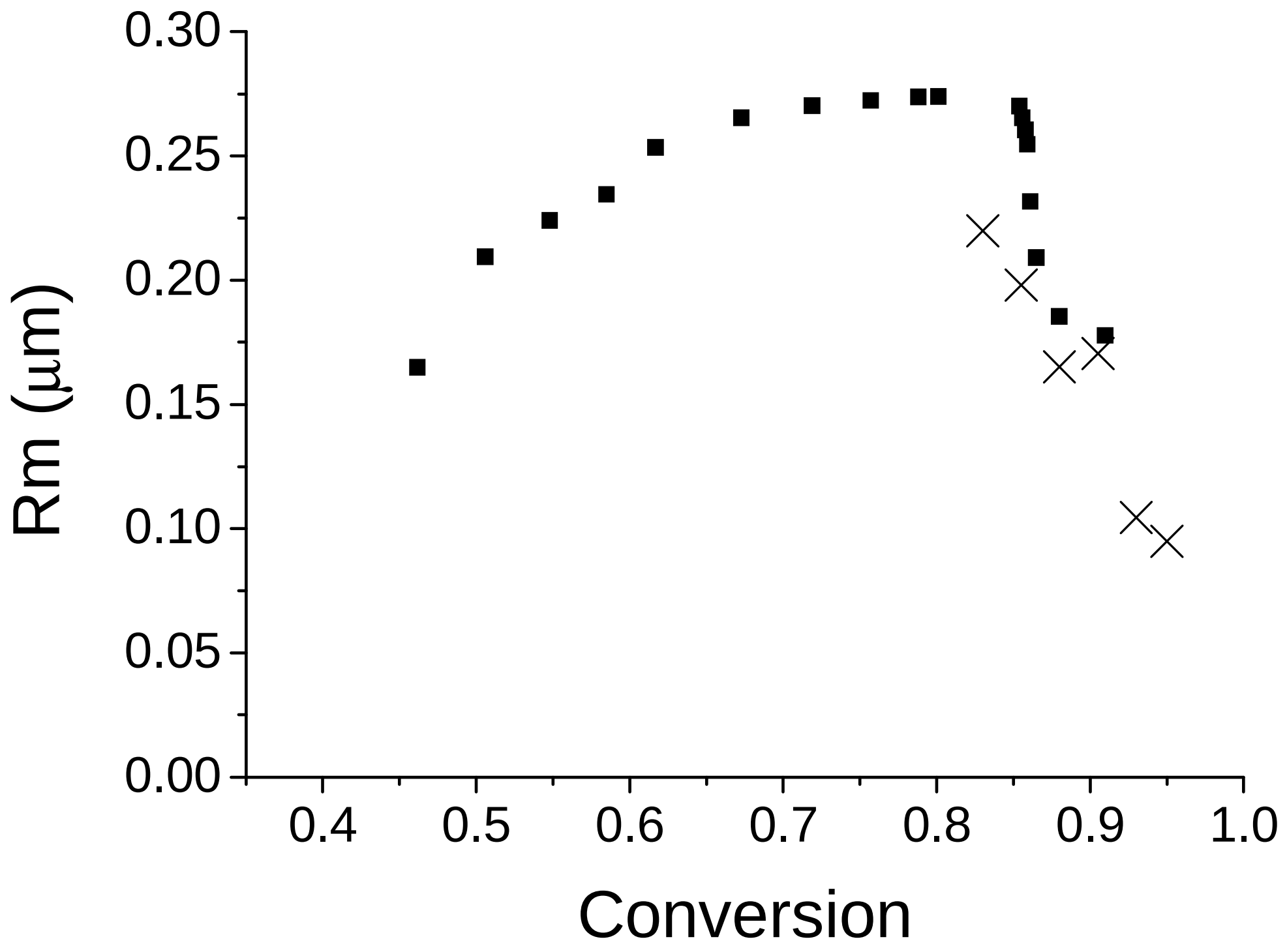

Fig. 2. Average radius of dispersed-phase droplets as a function of the monomer conversion (full symbols are values predicted from the fitting of LS curves using Pedersen's model; crosses are values determined in SEM images).

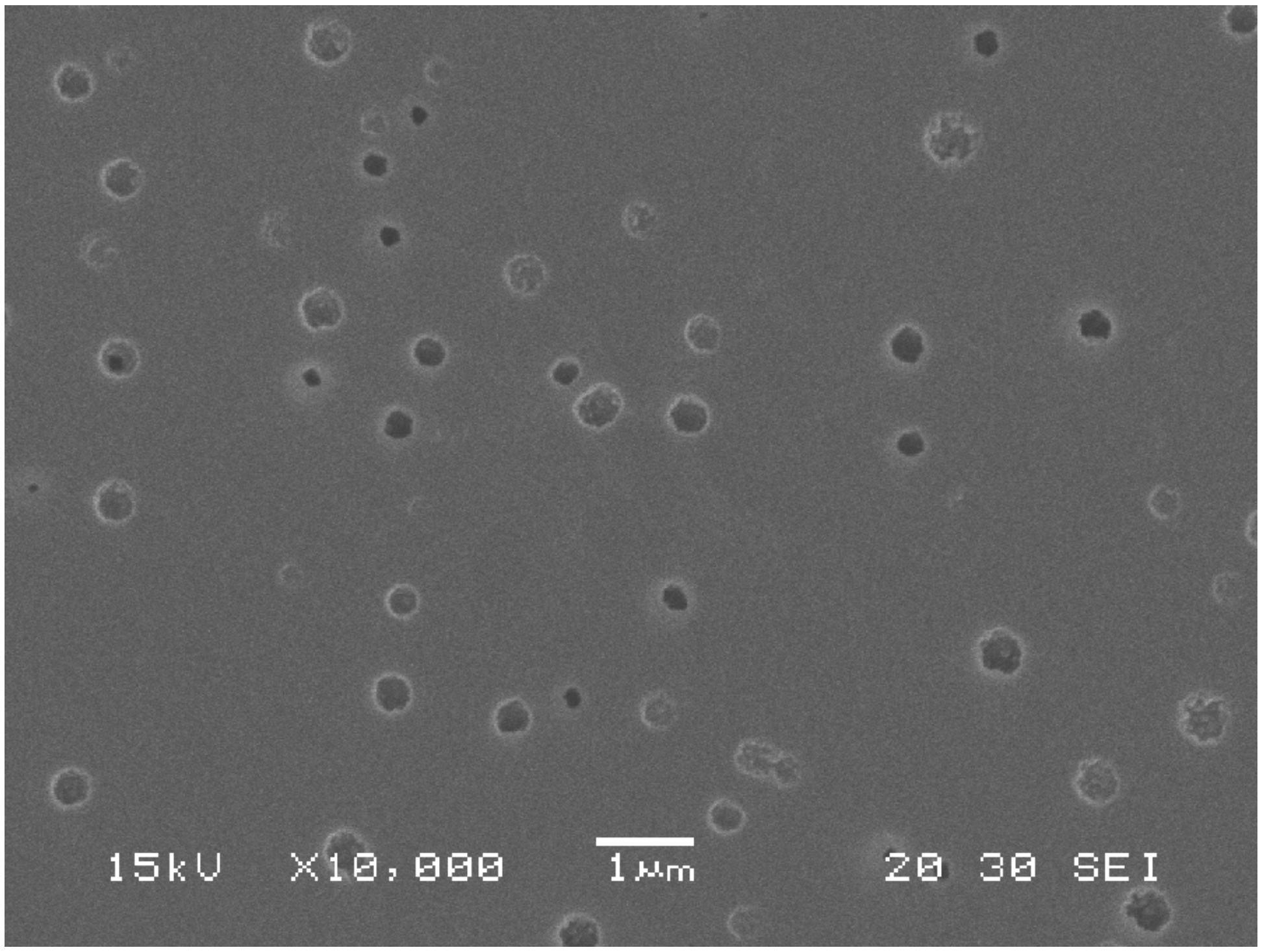

**(a)**

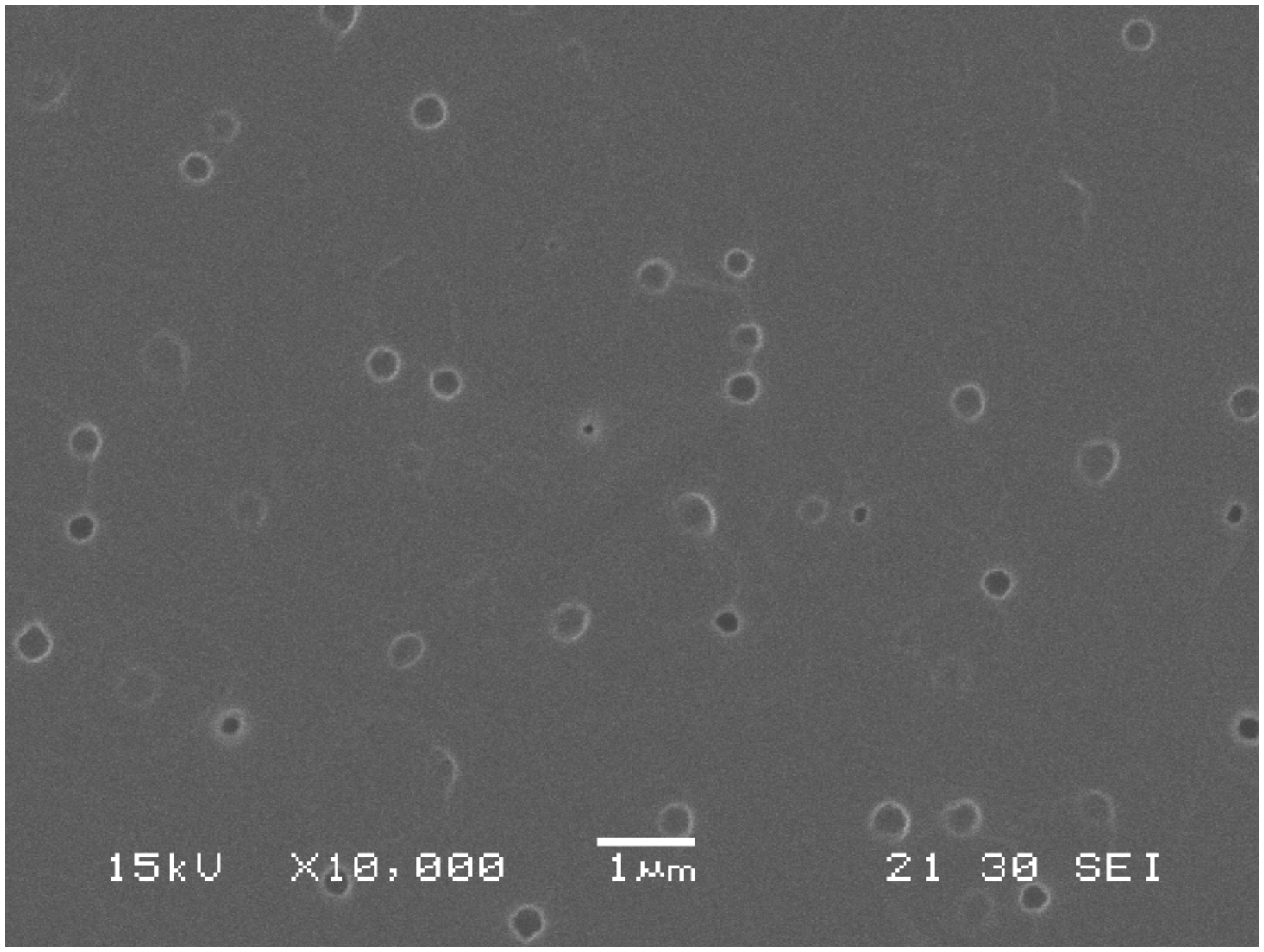

**(b)**

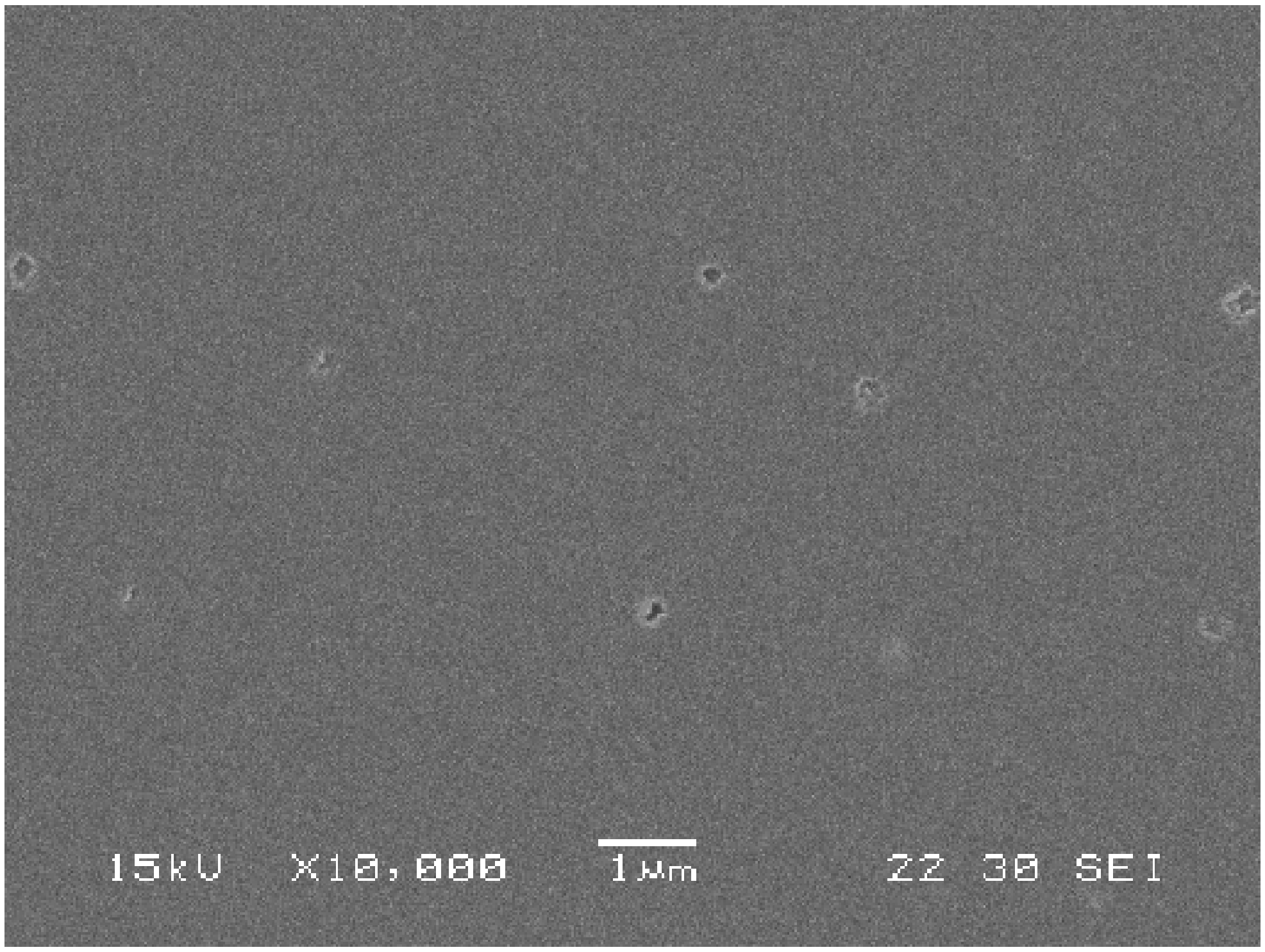

**(c)**

Fig. 3. SEM micrographs obtained at different monomer conversions; (a) 0.85, (b) 0.905, (c) 0.95.

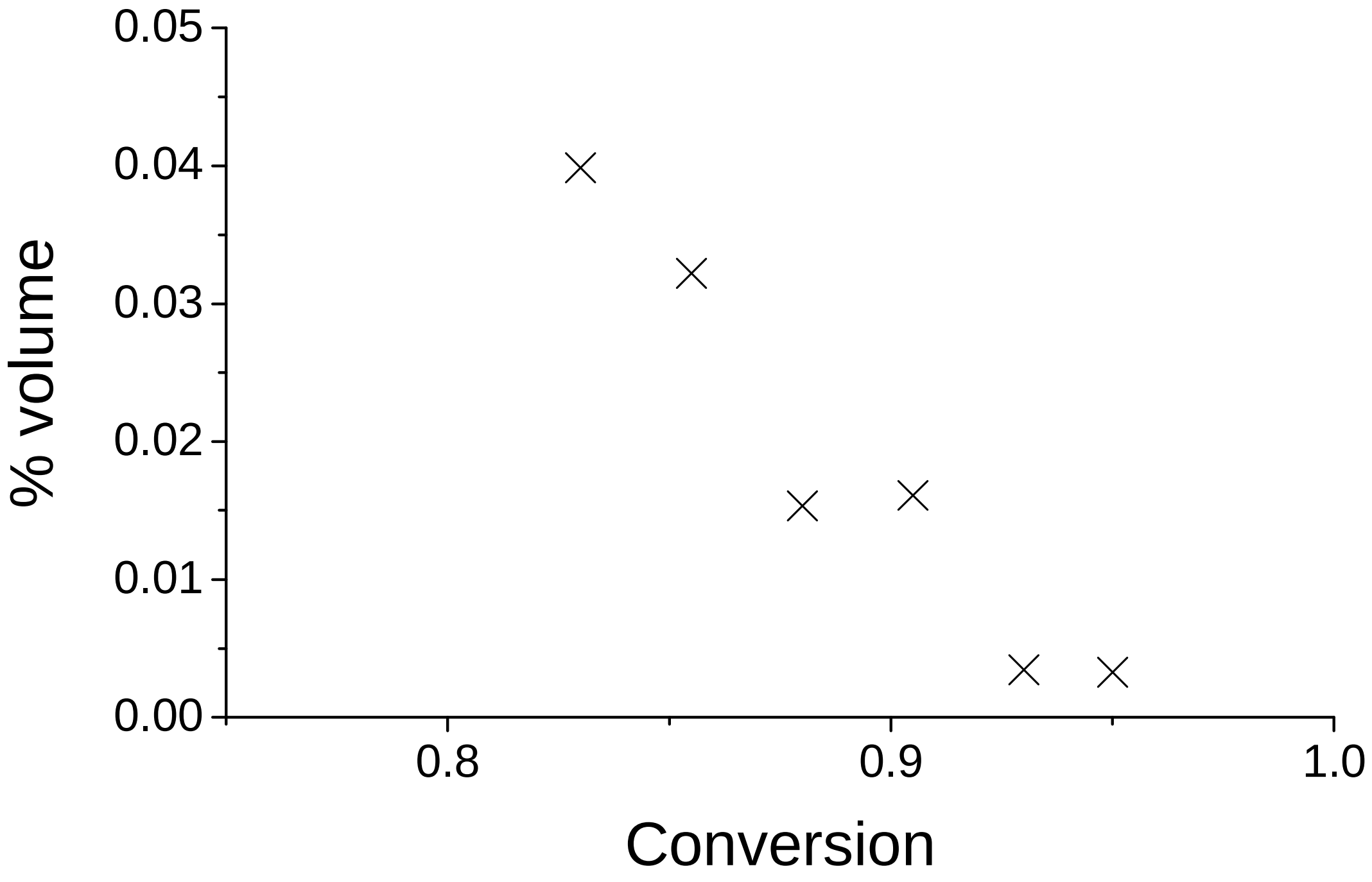

Fig. 4. Evolution of the volume fraction of dispersed phase as a function of conversion.

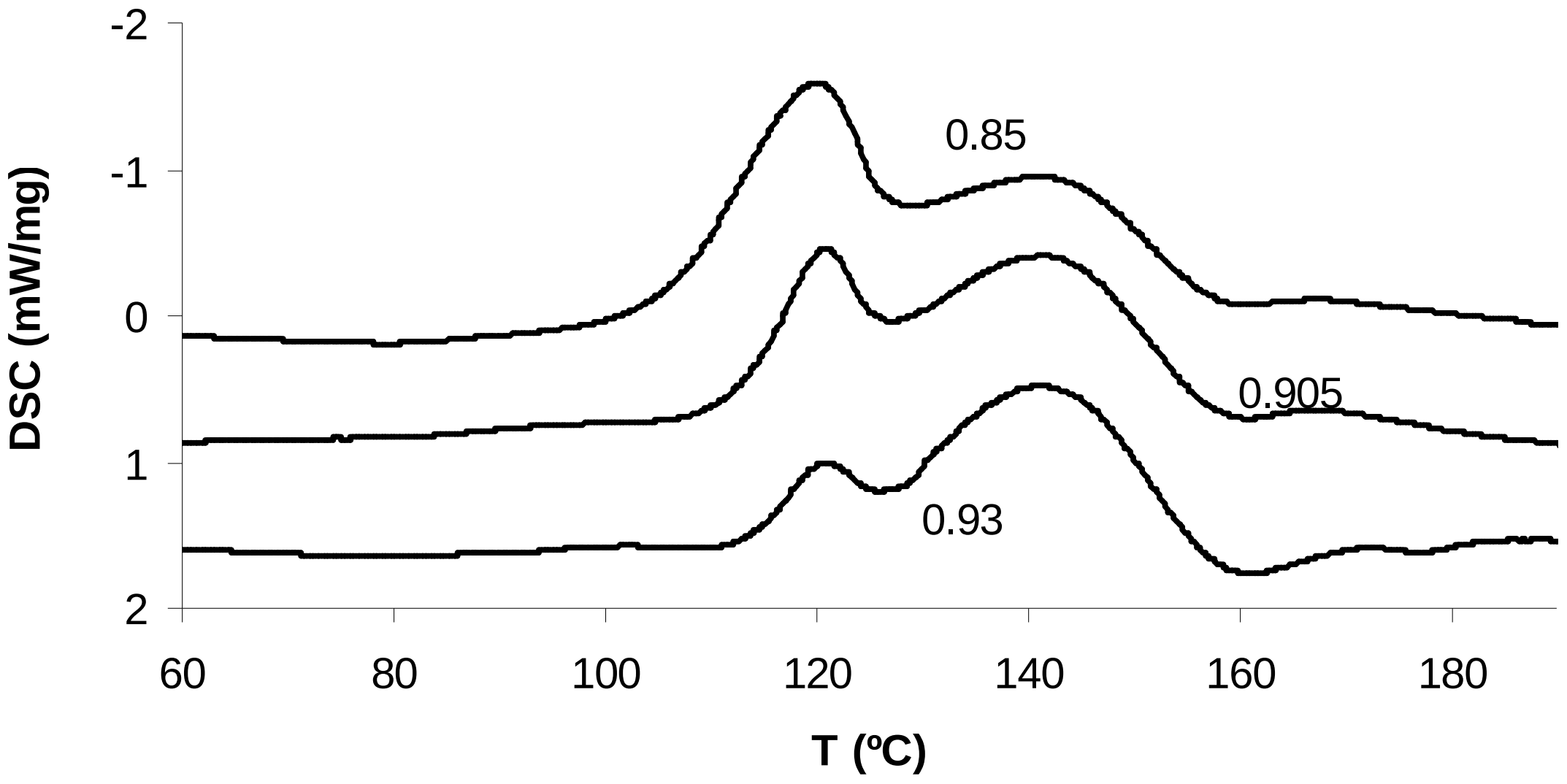

Fig. 5. Polymerization of the residual monomer as shown by DSC scans of samples previously reacted to the indicated conversions.

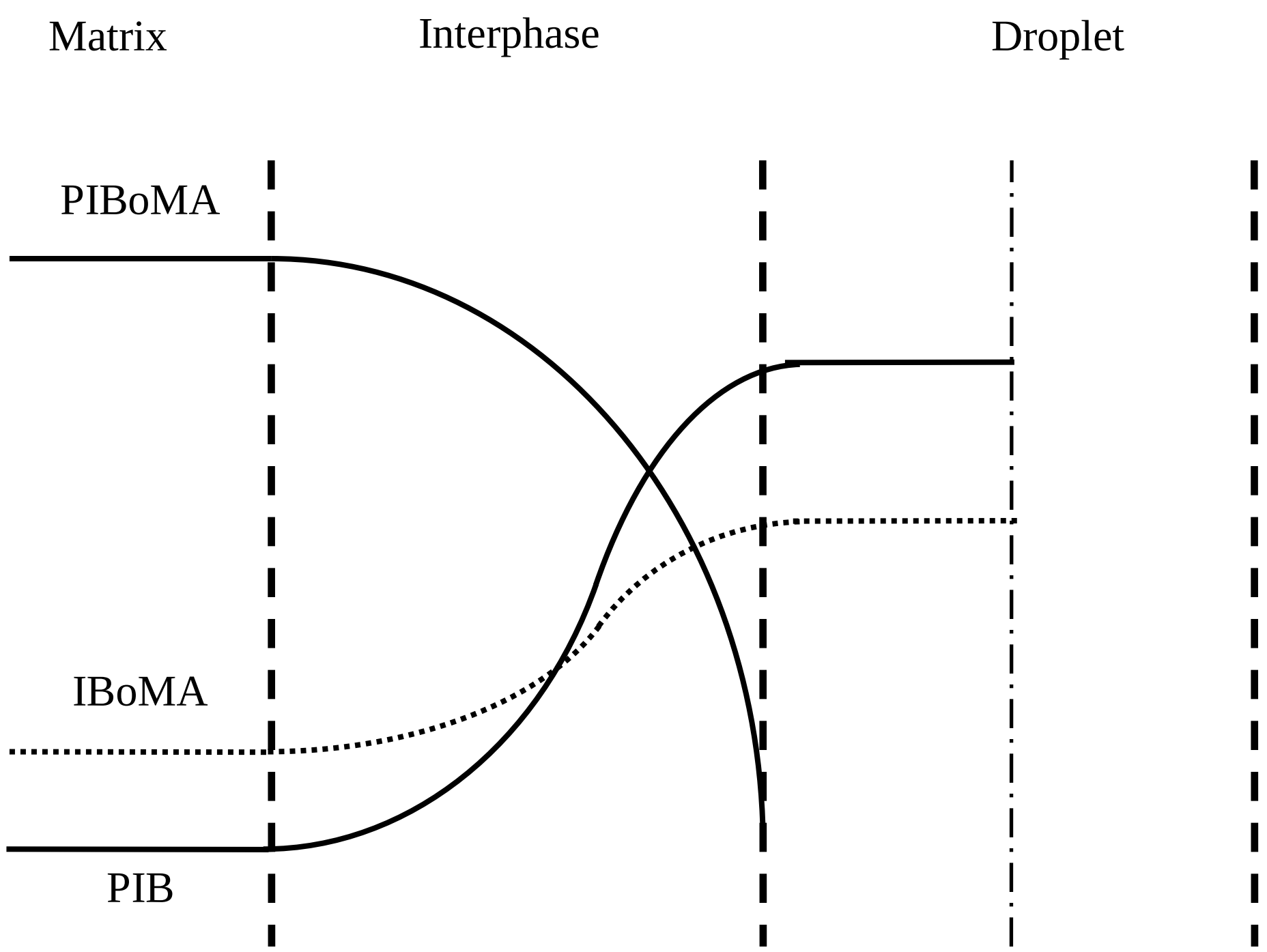

Fig. 6. Qualitative plot of concentration profiles in a region close to a droplet at the time when the matrix vitrifies.